\documentclass[conference]{IEEEtran}
\usepackage{graphicx, epstopdf}
\usepackage{epsfig,cite}
\usepackage{color}
\usepackage{amssymb,amsmath,mathtools, mathrsfs}
\usepackage{enumitem}
\usepackage{graphicx}
\usepackage{epstopdf}
\usepackage{caption}
\definecolor{blue}{rgb}{0,0,1}
\definecolor{darkgreen}{rgb}{0,.5,0}
\definecolor{darkred}{rgb}{.5,0,0}

\newtheorem{theorem}{Theorem}
\newtheorem{proposition}{Proposition}

\newtheorem{rem}{Remark}

\newtheorem{definition}{Definition}
\newtheorem{cor}{Corollary}

\newtheorem{proper}{Property}

\def\levy{L\'evy }

\newcommand{\realSet}{\mathcal{R}}

\def\Pr{{\mathrm{Pr}}}

\DeclareMathOperator\erfc{erfc}
\DeclareMathOperator\gsnr{G-SNR}
\newcommand{\thr}{\mathsf{th}}
\def\sgn{\mathop{\rm sgn}\nolimits} 
\def\StabDist{{\mathscr{S}}}

\IEEEoverridecommandlockouts

\begin{document}
\bstctlcite{ICC09_Ref2:BSTcontrol}

\title{On the Impact of Time-Synchronization in Molecular Timing Channels}

\author{ \IEEEauthorblockN{Nariman~Farsad, 
		Yonathan~Murin, 
		Weisi~Guo, 
	    Chan-Byoung~Chae,  
		Andrew~Eckford, 
		and Andrea Goldsmith} 
		\vspace{-2cm}

%
}

%
\maketitle
\vspace{-1cm}

\begin{abstract}
This work studies the impact of time-synchronization in molecular timing (MT) channels by analyzing three different modulation techniques. The first requires transmitter-receiver synchronization and is based on modulating information on the release timing of information particles. The other two are asynchronous and are based on modulating information on the relative time {\em between} two consecutive releases of information particles using {\em indistinguishable} or {\em distinguishable} particles. All modulation schemes result in a system that relate the transmitted and the received signals through an additive noise, which follows a stable distribution. As the common notion of the variance of a signal is not suitable for defining the power of stable distributed signals (due to infinite variance), we derive an expression for the geometric power of a large class of stable distributions, and then use this result to characterize the geometric signal-to-noise ratio (G-SNR) for each of the modulation techniques. In addition, for binary communication, we derive the optimal detection rules for each modulation technique. Numerical evaluations indicate that the bit error rate (BER) is constant for a given G-SNR, and the performance gain obtained by using synchronized communication is significant. Yet, it is also shown that by using two {\em distinguishable} particles per bit instead of one, the BER of the asynchronous technique can approach that of the synchronous one. 
\end{abstract}


\section{Introduction}
\vspace{-0.15cm}

Molecular communication is a biologically-inspired form of communication, where chemical signals are used to transfer information \cite{far16ST}. It is possible to modulate information on the particles using different techniques such as concentration, type, number, time of release, or a combination of those. Information particles can be transported from the transmitter to the receiver using diffusion, active transport,  bacteria, and flow (see \cite{far16ST} and the references therein). Among all these techniques, diffusion and flow-based propagation are the easiest to implement, and a few experimental platforms have been built to demonstrate molecular communication based on these transport mechanisms \cite{far13}.

In this work, we focus on modulation techniques for molecular communication and their corresponding system models.
Most prior modulation techniques rely on the concentration or the type of the released particles (e.g., \cite{nr13}). 
In this work, we consider molecular timing (MT) channels where timing-based modulation is employed. In biology, time of release may be used in the brain at the synaptic cleft, where two chemical synapses communicate over a chemical channel \cite{bor99}. Only a few works have considered this type of modulation: In \cite{eck09}, the time of release of the particles is used for encoding information, while in \cite{kri13} the information is encoded in the time interval between two pulse releases of information particles. Note that molecular timing channels can be different from queue-based timing channels \cite{ana96} since information particles can arrive out of order. 

In \cite{far15GLOBCOM} we proposed three general timing-based modulation techniques, which result in three different MT systems that relate the transmitted and received signals through additive noise. It was shown that for diffusion-based MT (DBMT) systems that rely on diffusive transport, the additive noise term falls in the stable distribution family~\cite{nol15}.  In particular, we consider an MT system, where information is encoded in the release timing of information particles (system A); modulation on the relative time between two consecutive releases of {\em indistinguishable} information particles (system B); and modulation on the relative time between two consecutive releases of {\em distinguishable} information particles (system C). The modulation scheme in system A requires synchronization between the transmitter and the receiver, while the modulation techniques in systems B and C are asynchronous. It must be noted that additive stable distributed noise has been used for modeling in a number of fields including modeling acoustic channels~\cite{he14}, and  ultra-wide bandwidth systems~\cite{nir09,fan12}. Therefore, the results of this paper could also be applicable in those areas. 

The main difference between the above modulation techniques is their time-synchronization requirements. Thus, by comparing the system performance of these modulation techniques, we gain knowledge about the significance of time-synchronization in MT systems. Note that it is possible to synchronize a transmitter with a receiver in molecular communication \cite{fel14,sha13CL}. There will be, however, an overhead for synchronization, which can be significant for diffusion-based propagation.   
In \cite{far15GLOBCOM} it is shown that the additive noise in the above three systems is heavy-tailed with infinite variance. Thus, using the signal variance as a measure for its power is not suitable. 
Instead, we derive expressions for the geometric power \cite{gon06} of almost all stable distributions, and use it to represent the noise power. Furthermore, instead of using the well known signal-to-noise ratio (SNR) metric, we use the geometric SNR (G-SNR)~\cite{gon06} metric, which is given by the power of the input signal divided by the geometric power of the noise with some normalization constants. 
To compare the performance of the three proposed modulation techniques, we consider a binary communication system and derive the optimal detection rule for each modulation technique. We then show numerically that the bit error rate (BER) is constant for a given G-SNR regardless of the input signal power and the geometric noise power, which supports the fact that G-SNR correctly captures the quality of heavy-tailed additive noise in general, and of additive noise in MT channels in particular. Note that it is very challenging to show this analytically because the probability density function (PDF) of the additive noise cannot be represented in terms of elementary functions. 

Finally, we numerically compare the BER of all three modulation techniques. Again note that closed-form expression are difficult to derive because of complex PDF expressions. We show that system B with indistinguishable particles exhibits the highest BER, while system A achieves the lowest BER. This demonstrates the significant performance gain that can be achieved in time-synchronized transmission over MT channels. 
We further show that by adjusting the diffusion coefficients of the information particles in system C, which assumes asynchronous transmission, its BER can approach the BER of system A, in which synchronization is assumed. 
However, this comes at the cost of using two distinguishable information particles per bit instead of a single information particle per bit. 
Although this cost has been captured in the G-SNR (i.e., at a given separation distance the G-SNR of system C can be four times that of system A), both the transmitter and receiver must be capable of transmitting and detecting two distinguishable particles.  

The rest of this paper is organized as follows. In Section \ref{sec:model} we present the three timing-based modulation techniques, and derive an additive noise system model for each of them. The noise geometric power and the G-SNR for each system is derived in Section \ref{sec:SNR}.  In Section \ref{sec:PBE}, binary communication is studied, and the optimal detectors are derived. Numerical BER evaluations of the proposed modulation techniques are presented in Section \ref{sec:numEval}, and Section \ref{sec:conc} concludes the paper.   

\textbf{\textit{Notation:}} We denote the set of real numbers by $\realSet$, and the set of positive real numbers by $\realSet^{+}$. 
We denote random variables (RV)s with upper case letters, e.g., $X$, and their realizations with the corresponding lower case letters, e.g., $x$. We use $f_{Y}(y)$ to denote the PDF of a continuous RV $Y$ on $\realSet$, $f_{Y|X}(y|x)$ to denote the conditional PDF, and  $F_{Y}(y)$ and $F_{Y|X}(y|x)$ to denote the corresponding cumulative distribution functions (CDF). 
Finally, we use $\Re\{ z \}$ to denote the real part of the complex number $z$, and $\erfc\left( \cdot \right)$ to denote the complementary error function given by $\erfc(x) = \frac{2}{\sqrt{\pi}} \int_{x}^{\infty}{e^{-u^2} du}$.

\section{The Modulation Techniques}
\label{sec:model}
\vspace{-0.15cm}

We begin this section with several general assumptions for communication over MT channels.

\subsection{Communication over MT Channels - General Assumptions}
\vspace{-0.15cm}
Throughout this paper, we assume that the transmitter perfectly controls the release time of each information particle, while the receiver perfectly measures the arrival times of the information particles. 
	Furthermore, the transmitter and the receiver are assumed to be perfectly synchronized in time, {\em when synchronization is required by the modulation scheme}, i.e., in system A.	
	Information particles that arrive at the receiver are assumed to be absorbed and hence removed from the propagation medium.
	All information particles are assumed to propagate independently of each other, and their trajectories are random according to an independent and identically distributed (i.i.d.)  random process. 
	Finally, it is assumed that there is no inter-symbol interference (ISI) between consecutive channel uses.\footnote{In \cite{nanocom16} we study communication over DBMT channels in the presence of ISI.} In practice this assumption can be satisfied if the time between consecutive channel uses is large enough or if chemical reactions are used to dissipate the particles \cite{noe14}.
 Note that these assumptions have also been used in many previous works.

As the additive noises considered in this work follow stable distributions, we next define this class of distributions and briefly discuss their properties.

\subsection{Stable Distributions}
\vspace{-0.15cm}
Let $X_1$ and $X_2$ be independent copies of a random variable $X$. Then $X$ is stable distributed if, for any constants $a > 0$ and $b > 0$, the random variable $aX_1 + bX_2$ has the same distribution as $cX + d$, for some constants $c > 0$ and $d$. Stable distributions can be defined via their characteristic function as follows.
\begin{definition}
	Let $-\infty < \mu < \infty, c\ge 0, 0 < \alpha \le 2$, and $-1 \le \beta \le 1$. Further define: 
	\vspace{-0.15cm}
	\begin{align*}
	\Phi(t,\alpha) \triangleq \begin{cases} 
	\tan \left( \frac{\pi \alpha}{2}\right), & \alpha \ne 1 \\ 
	-\frac{2}{\pi} \log (|t|), & \alpha = 1
	\end{cases}.
	\end{align*}
	
	\vspace{-0.1cm}
	\noindent Then, the characteristic function of a stable RV $X$, with location parameter $\mu$, scale parameter $c$, characteristic exponent $\alpha$, and skewness parameter $\beta$, is given by:
	\vspace{-0.15cm}
	\begin{align*}
	\varphi (t;\mu,c,\alpha,\beta) \mspace{-3mu} = \mspace{-3mu} \exp \left[ j \mu t \mspace{-3mu} - \mspace{-3mu}| c t |^\alpha (1 \mspace{-3mu} - \mspace{-3mu} j\beta \sgn(t) \Phi(t,\alpha) ) \right].
	\end{align*}
\end{definition} 

\vspace{-0.1cm}
In the following, we use the notation $\mathscr{S}(\mu, c, \alpha, \beta)$ to represent a stable distribution with the parameters $\mu, c, \alpha$, and $\beta$. One interesting property of stable distributions is the following.
\begin{proper}
	\label{prop:standStable}	
	Let $\tilde{X} \sim \StabDist(0,1,\alpha,\beta)$ be the standard stable RV with parameters $\alpha$ and $\beta$. Then the PDF and the CDF of any RV $X \sim \StabDist(\mu,c,\alpha,\beta)$ can be calculated as
	\vspace{-0.15cm}
	\begin{align*}
	f_X(x) = \frac{f_{\tilde{X}}\big( \tfrac{x-\mu}{c} \big)}{c}, \qquad F_X(x) =  F_{\tilde{X}}( \tfrac{x-\mu}{c} ). 
	\end{align*}
\end{proper}

\vspace{-0.1cm}
Next, we discuss the three modulation techniques presented in \cite{far15GLOBCOM}, and discuss the resulting system models.

\subsection{System A}
\vspace{-0.15cm}
The first MT modulation technique is the one proposed in \cite{eck09}, where the information is encoded in the release timing of a single information particle. 
Let $T_{x} \in \mathcal{T}\subseteq\realSet^+$ be the release time of the information particle at the transmitter. In this scheme, the information is modulated onto the release time itself. The released particle is then randomly transported from the transmitter to the receiver. 
Letting $T_{y}$ be the time of arrival at the receiver, we write:
\vspace{-0.15cm}
\begin{align}
\label{eq:timingChA}
	T_{y} = T_{x} + T_n,
\end{align}

\vspace{-0.1cm}
\noindent where $T_n$ is the random propagation delay of the information particle. We refer to \eqref{eq:timingChA} as system A. One of the main challenges of this modulation scheme is the need for synchronization between the transmitter and the receiver. In this work, whenever system A is used we assume the transmitter and the receiver are perfectly synchronized. In \cite{far15GLOBCOM} it was shown that for diffusion-based transport in this synchronized case, the additive noise $T_n$ is a stable distributed random variable.  

Let $d$ be the distance between the transmitter and the receiver, and $D$ be the diffusion coefficient of the particle. For diffusion-based transport, we have $T_n \sim \StabDist (0,c_A, \tfrac{1}{2},1)$, where $c_A \triangleq \tfrac{d^2}{2D}$. Note that this subclass of stable distributions is known as the L\'evy distribution, and its PDF is given in \cite[Eq. (1)]{far15GLOBCOM}.

To avoid the synchronization requirement in system A, we now discuss two other modulation schemes in which information is modulated on the time duration {\em between} two consecutive releases of information particles. The receiver decodes the information from the time between the arrivals of two molecules. Note that in this case synchronization between the transmitter and the receiver is not required.

\subsection{System B}
\vspace{-0.15cm}
This modulation technique uses two {\em indistinguishable} information particles, i.e., the information particles are of the same type. Without loss of generality, let $T_{x_1}$ be the release timing of the first information particle and $T_{x_2}$ be the release timing for the second information particle, with $T_{x_2}>T_{x_1}$. The information is encoded in $L_x = T_{x_2}-T_{x_1}$. Using \eqref{eq:timingChA}, the model for this modulation scheme is given by:
\vspace{-0.15cm}
\begin{align}
\label{eq:timingChB}
	|T_{y_2} - T_{y_1}|&= |T_{x_2}-T_{x_1} + T_{n_2} - T_{n_1}|,  \nonumber \\
	L_y &= | L_x+L_n|,
\end{align}

\vspace{-0.1cm}
\noindent where $L_n = T_{n_2} - T_{n_1}$ is the random noise and $T_{n_2}$ and $T_{n_1}$ are independent and identically distributed noise terms as in \eqref{eq:timingChA}. We refer to \eqref{eq:timingChB} as system B. Note that the absolute value in this formulation is due to the fact that both information particles are indistinguishable, and therefore the receiver can observe only the absolute difference of arrival times. In \cite[Theorem 1]{far15GLOBCOM}, it was shown that for diffusion-based transport, the noise term $L_n$ is distributed according to $L_n \sim \StabDist (0,c_B, \tfrac{1}{2},0)$, where $c_B\triangleq \frac{2d^2}{D}$. The PDF of $L_n$ is given in \cite[Eq. (19)]{far15GLOBCOM}.

\subsection{System C}
\vspace{-0.15cm}
The last modulation scheme encodes information in the time between releases of two {\em distinguishable} information particles (i.e., two different particle types) to encode information. Let $T_{x_a}$ be the release timing of the type-$a$ information particle and $T_{x_b}$ be the release timing for the type-$b$ information particle. We assume the information is encoded in $Z_x = T_{x_b}-T_{x_a}$. Unlike \eqref{eq:timingChB} where $L_x$ is always positive, $Z_x$ can be positive or negative depending on the order that the type-$a$ and type-$b$ information particles are released. Using \eqref{eq:timingChA}, the model for this scheme is given by:
\vspace{-0.15cm}
\begin{align}
\label{eq:timingChC}
	T_{y_b} - T_{y_a} &= T_{x_b}-T_{x_a} + T_{n_b} - T_{n_a},  \nonumber \\
	Z_y &= Z_x+ Z_n, 
\end{align}

\vspace{-0.1cm}
\noindent where $Z_n = T_{n_b} - T_{n_a}$ is the random additive noise and $T_{n_b}$ and $T_{n_a}$ are independent noise terms as in (\ref{eq:timingChA}). We refer to \eqref{eq:timingChC} as system C. Again, no synchronization is required between the transmitter and receiver.  In \cite[Theorem 2]{far15GLOBCOM}, it was shown that for diffusion-based transport, the noise term $Z_n$ is a stable distributed random variable. Let $D_a$ be the diffusion coefficient of the type-$a$ particle and $D_b$ be the diffusion coefficient of the type-$b$ particle. The additive noise term is stable distributed with $Z_n \sim \StabDist (0,c_C, \tfrac{1}{2},\beta_C)$ where $c_C \triangleq \frac{d^2(\sqrt{D_a}+\sqrt{D_b})^2}{2D_aD_b}$,  $\beta_C \triangleq  \frac{\sqrt{D_a}-\sqrt{D_b}}{\sqrt{D_a}+\sqrt{D_b}}$. The PDF of $Z_n$ was derived in \cite[Eq. (25)]{far15GLOBCOM}.

In the next section we define the strength of the noise in each model using the geometric power framework \cite{gon06}.

\section{Geometric Power and G-SNR} \label{sec:SNR}
\vspace{-0.15cm}

We first note that all stable distributions, apart from the case $\alpha = 2$, have infinite variance. 
In fact, this statement can be generalized to moments of order $p \le \alpha$, see \cite{gon06}. Therefore, the conventional notion of power, which is based on the variance of a signal, is not informative in the case of stable RVs with $\alpha < 2$. 
In this section we use a more generalized definition of power, the {\em geometric power}, as proposed in \cite[Section III]{gon06}. This definition uses zero-order statistics, i.e., it is based on logarithmic ``moments" of the form $\mathbb{E}[\log|T_n|]$.

\begin{definition}
	The geometric power of the RV $N$ is given~by:
	\vspace{-0.2cm}
	\begin{align*}
	S_0(T_n) \triangleq e^{\mathbb{E} [\log|N|]}.
	\end{align*}  
\end{definition}

\vspace{-0.1cm}
In \cite[Prop. 1]{gon06}, an expression for the geometric power of a {\em symmetric} stable distribution (i.e., $\StabDist (0,c,\alpha,0)$) is presented. This expression can therefore be used to calculate the geometric power of the noise term $L_N$ in system B. Yet, this expression is {\em not applicable} for the noise terms in systems A and C in which $\beta \neq 0$. 
In the following theorem we present the geometric power for almost all stable distributions.
\begin{theorem}
	\label{thrm:GpowerStable}
	Let $N \sim \StabDist (0,c,\alpha, \beta)$, where $\alpha \neq 1$, or $\alpha = 1$ and $\beta=0$. Then, the geometric power of $N$ is given by:
	\begin{align*}
		S_0(N) = c G_\gamma^{(1/\alpha-1)} \big(1+\beta^2\tan^2(\tfrac{\pi \alpha}{2})\big)^{1/(2\alpha)}, 
	\end{align*}
	where $G_\gamma = e^\gamma$, and $\gamma \approx 0.5772$ is the Euler's constant \cite[Ch. 5.2]{nist10}.
\end{theorem}
\begin{IEEEproof}
	The proof is provided in \cite{BerMTExtended}.
\end{IEEEproof}

	For the systems considered in this paper, since $\alpha = \frac{1}{2}$, the noise power simplifies to: 
	\begin{align*}
		S_0(N) = cG_\gamma\big(1+\beta^2\big).
	\end{align*}
	Note that in this case, the noise power increases with respect to $\beta$ (the degree of skewness) and $c$ (the scale parameter).
Using this definition of geometric power, we define the geometric SNR (G-SNR) as in \cite[Section III]{gon06}:
\begin{definition}
	\label{def:GSNR}
	Let $X$ be the input signal in an additive-noise channel with a random noise $N$. Then the G-SNR is defined~as:
	\begin{align}
	\label{eq:GSNRdef}
	\gsnr \triangleq \frac{1}{2 G_\gamma }\bigg(\frac{X_{\max} - X_{\min}}{S_0(N)}\bigg)^2,
	\end{align}  
	where $X_{\max}$ and $X_{\min}$ are the maximum and minimum admissible values for the input $X$. The normalizing term $\frac{1}{2 G_\gamma}$ is used to ensure that the G-SNR corresponds to the standard SNR in the case of an additive Gaussian noise.
\end{definition}

The following corollary states the G-SNR of systems A and C.
\begin{cor}
Let $\Delta$ be the time period over which a transmission symbol could be encoded. Then, the G-SNR of systems A and C is given by:
\begin{align}
	\gsnr_A &= \frac{1}{2 G_\gamma }\bigg(\frac{\Delta}{2 c_A G_\gamma }\bigg)^2, \label{eq:GSNRa}\\
	\gsnr_C &= \frac{1}{2 G_\gamma }\bigg(\frac{2\Delta}{ c_C G_\gamma(1+\beta_C^2) }\bigg)^2. \label{eq:GSNRc}
	\end{align} 	
\end{cor}
	
\begin{rem} \label{rem:absValImpact}
	Note that system B involves an absolute value operation, thus, the G-SNR of system B cannot be obtained using the techniques used to derive the G-SNR for systems A and C. Since the absolute value operation is non-invertible, the G-SNR of $L_y = L_x + L_n$ can serve as an upper bound on the G-SNR of system B. This upper bound is given by:
	\vspace{-0.15cm}
	\begin{align}
		\gsnr_B \le \gsnr_B^{\text{ub}} = \frac{1}{2 G_\gamma }\bigg(\frac{\Delta}{ c_B G_\gamma }\bigg)^2. \label{eq:GSNRb}
	\end{align}
	
	\vspace{-0.1cm}
\noindent Our numerical evaluations in Section \ref{sec:numEval} indeed indicate that the BER for the optimal detector for system B is higher than the BER of the optimal detector for $L_y = L_x + L_n$, which can be obtained from system C by using $D_a = D_b$.
\end{rem}

	When the diffusion coefficient and the distance between the transmitter and the receiver are the same, the G-SNR of system A is at least four times larger than the G-SNR of system B (i.e., $c_B=4c_A$). 
	This implies that although two information particles are released in system B, while only a single particle is released in system A, the gain from synchronization is a factor of at least $4$ in the noise geometric power.
%
\begin{rem}
\label{rem:effectBeta}
For system C let $r=D_a/D_b$ be the ratio of the diffusion coefficient of the two information particles. Then the noise parameters can be written as $c_C= \frac{d^2(\sqrt{r}+1)^2}{2rD_b}$ and $\beta_C = \frac{\sqrt{r}-1}{\sqrt{r}+1}$. Next, assume that the diffusion coefficient $D_b$ is fixed, and the diffusion coefficient of $D_a$ can be changed. In this case the noise geometric power is proportional to $\frac{1}{r}$, which decreases as $r$ increases. This also implies that the  G-SNR increases with $r$. From the expression for $\beta_C$ and $c_C$ we observe that when $r \rightarrow \infty$, then $\beta_c \rightarrow 1$ and $c_C \rightarrow \tfrac{d^2}{2D_b}$. Thus, in this case system C reduces to system A, while no synchronization is required between the transmitter and the receiver.
Yet, this comes at a cost of using two different information particles. This cost is partly captured in the G-SNR expression since the geometric power of the transmitted signal in system C is four times that of systems A and B.
\end{rem}

Next, we discuss optimal detection in binary transmission over systems A, B, and C.

\section{Optimal Detection in Binary Communication}
\label{sec:PBE}
\vspace{-0.1cm}
We consider equiprobable binary transmission over the three different DBMT systems. Using the PDFs of the noise models developed in \cite{far15GLOBCOM}, we characterize the optimal detection rule for each modulation scheme. The BER for each system is evaluated numerically in Section \ref{sec:numEval} illustrating the gain from synchronization. Note that it is very difficult to derive closed-form expressions for the BER because the PDF expressions are very complex. We begin with system A.

\vspace{-0.1cm}
\subsection{System A}
\vspace{-0.15cm}
Let $T_x \in \{0,\Delta\}$, where $\Delta>0$. 
Using Property \ref{prop:standStable}, we write the distribution of the system output, conditioned on the system input, in terms of the standard \levy distribution $\tilde{T}_n \sim \StabDist(0,1, \tfrac{1}{2},1)$ as follows:
\vspace{-0.15cm}
\begin{align*}
f_{T_y|T_x}(t_y|T_x=0) &= \frac{f_{\tilde{T}_n}(t_y/c_A)}{c_A}
\end{align*}
\vspace{-0.4cm}
\begin{align*}
f_{T_y|T_x}(t_y|T_x=\Delta) &= \frac{f_{\tilde{T}_n}\big((t_y-\Delta)/c_A\big)}{c_A}. 
\end{align*}

\vspace{-0.1cm}
As the two transmitted symbols are equiprobable, the detector that minimizes the probability of error is the maximum likelihood (ML) detector. 
The likelihood ratio for this detection problem is given by:
\vspace{-0.15cm}
\begin{align}
\Lambda_A (t_y) = \frac{f_{T_y|T_x}(t_y|T_x=0)}{f_{T_y|T_x}(t_y|T_x=\Delta)}, \label{eq:LR_A}
\end{align}

\vspace{-0.1cm}
\noindent and optimal detection can be done by a comparison of the log likelihood ratio (LLR) to zero:
\vspace{-0.2cm}
\begin{align*}
\log(\Lambda_A (t_y) ) \begin{matrix} T_x = 0 \\ \gtrless \\ T_x = \Delta \end{matrix} 0.
\end{align*} 

\vspace{-0.1cm}
\noindent Note that the proof of the existence of the optimal threshold value is straightforward using the fact that stable distributions are unimodal \cite[Theorem 2.7.6]{zol86-book}, and that for the noise term $T_n$ the mode is at $c/3$. Therefore, there exists a $\Delta<\thr_A\leq c/3+\Delta$, such that $\Lambda_A (t)>1$ for $t<\thr_A$  and  $\Lambda_A (t)\leq1$ for $t\geq\thr_A$\footnote{The optimal threshold and the BER are analytically derived in another work \cite{mur16}. Note that this derivation is possible for the L\'evy noise because it has a known PDF, and would be very challenging task for Systems B and C.}. 

The probability of error for the modulation scheme in system A is now given by:
\vspace{-0.15cm}
\begin{align*}
P_e^A &= P(T_x=0) \Pr(t_y>\thr_A | T_x = 0) \nonumber \\
&\qquad +P(T_x=\Delta) \Pr(t_y\leq\thr_A | T_x= \Delta), \nonumber \\
&= 0.5 \Pr(t_n>\thr_A)+0.5 \Pr(t_n\leq\thr_A-\Delta) \nonumber \\
&= 0.5[1-F_{\tilde{T}_n}(\tfrac{\thr_A}{c_A})+F_{\tilde{T}_n}(\tfrac{\thr_A-\Delta}{c_A})],
\end{align*}

\noindent where $F_{\tilde{T}_n}(t)$ is the CDF of a standard L\'evy RV.

\vspace{-0.15cm}
\subsection{System B}

Let $L_x \mspace{-3mu} \in \mspace{-3mu} \{0,\Delta\}$, where $L_x \mspace{-3mu} = \mspace{-3mu} 0$ represents two particles released simultaneously, while $L_x=\Delta$ represents two particles released $\Delta$ seconds apart. Let $\tilde{L}_n \mspace{-3mu} \sim \mspace{-3mu} \StabDist(0,1,\tfrac{1}{2},0)$ be the standard form of the noise term in \eqref{eq:timingChB}.
First, we derive the PDF of the system output $L_y$, given the system input $L_x$, in terms of the PDF of $\tilde{L}_n$. 
\begin{proposition}
	The system output $L_y$, given the system input $L_x$, has the PDF: 
\vspace{-0.15cm}
\begin{align*}
f_{L_y|L_x}(\ell_y|L_x \mspace{-3mu} = \mspace{-3mu} 0) & \mspace{-3mu} = \mspace{-3mu}
\begin{cases}
\frac{2f_{\tilde{L}_n}\big(\tfrac{\ell_y}{c_B}\big)}{c_B} & \ell_y>0 \\
\frac{2}{(c_B\pi)} & \ell_y=0 \\
0 & \ell_y<0
\end{cases}, \\ 
f_{L_y|L_x}(\ell_y|L_x \mspace{-3mu} = \mspace{-3mu} \Delta) & \mspace{-3mu} = \mspace{-3mu}
\begin{cases}
\frac{f_{\tilde{L}_n}\big( \tfrac{\ell_y-\Delta}{c_B}\big)+f_{\tilde{L}_n}\big( \tfrac{-\ell_y-\Delta}{c_B}\big)}{c_B} & \ell_y>0 \\
\frac{f_{\tilde{L}_n}\big( \tfrac{\Delta}{c_B}\big)}{c_B} & \ell_y=0 \\
0 & \ell_y<0
\end{cases}. 
\end{align*}

\end{proposition}

\begin{IEEEproof}
	It is clear from the system definition that when $L_y<0$ the PDF is 0 (i.e. the time between two arrival times is not negative). When $L_y=0$, we have $f_{L_y|L_x}(0|L_x=0)=f_{L_n}(0)$, and $f_{L_y|L_x}(0|L_x=\Delta)=f_{L_n}(\Delta)$.  To derive the PDF value for $L_y>0$, we use the fact that the CDF of $L_y$, given $L_x=x\geq0$, can be obtained from the CDF of $\tilde{L}_n$ as:
	\vspace{-0.15cm}
	\begin{align*}
	F_{L_y|L_x}(\ell_y|L_x=x) &= \Pr(L_y \leq \ell_y|L_x=x), \\
	&= \Pr(|x+L_n| \leq \ell_y) \\
	&= \Pr(-\ell_y \leq x+L_n \leq \ell_y) \\
	&= \Pr(\tfrac{-\ell_y-x}{c_B} \leq \tilde{L}_n \leq \tfrac{\ell_y-x}{c_B} ) \\
	&= F_{\tilde{L}_n}\big(\tfrac{\ell_y-x}{c_B}\big)-F_{\tilde{L}_n}\big(\tfrac{-\ell_y-x}{c_B}\big)
	\end{align*}

	\vspace{-0.1cm}
	\noindent By differentiating with respect to $\ell_y$, and setting $x=0$ and $x=\Delta$ we obtain the desired PDFs.
\end{IEEEproof}

Similarly to \eqref{eq:LR_A}, the likelihood ratio for the ML detector for system B is given by:
\vspace{-0.15cm}
\begin{align*}
\Lambda_B (\ell_y) = \frac{f_{L_y|L_x}(\ell_y|L_x=0)}{f_{L_y|L_x}(\ell_y|L_x=\Delta)}.
\end{align*}

\vspace{-0.1cm}
\noindent The following theorem states that just like in the case of system A, the ML detector can be implemented by comparing $\log (\Lambda_B (\ell_y))$ to zero:
\begin{theorem}
	\label{thrm:ChanBthrExistance}
	There exists a fixed threshold $\thr_B >\tfrac{\Delta}{2}$ such that the ML detector in the case of system B is given by:
	\vspace{-0.15cm}
	\begin{align*}
		\log(\Lambda_B (\ell_y) ) \begin{matrix} T_x = 0 \\ \gtrless \\ T_x = \Delta \end{matrix} 0.
	\end{align*} 
\end{theorem}
\begin{IEEEproof}
	The proof is provided in \cite{BerMTExtended}.
\end{IEEEproof}

Finally, the probability of error for binary communication over system B is given by:
\vspace{-0.15cm}
\begin{align*}
P_e^B &= P(L_x=0) \Pr(L_y>\thr_B | L_x = 0) \nonumber \\
&\qquad +P(L_x=\Delta) \Pr(L_y\leq\thr_B | L_x= \Delta), \nonumber \\
&= 0.5(\Pr(L_n>\thr_B)+\Pr(L_n\leq -\thr_B) )\nonumber \\
&\qquad +0.5\Pr(-\thr_B-\Delta \leq L_n\leq \thr_B-\Delta), \nonumber \\
&= 0.5(\Pr(\tilde{L}_n>\tfrac{\thr_B}{c_B})+\Pr(\tilde{L}_n\leq -\tfrac{\thr_B}{c_B}) )\nonumber \\
&\qquad +0.5\Pr(\tfrac{-\thr_B-\Delta}{c_B} \leq \tilde{L}_n\leq \tfrac{\thr_B-\Delta}{c_B}), \nonumber \\
&= F_{\tilde{L}_n}\big(\tfrac{\thr_B}{c_B}\big) \mspace{-2mu} + \mspace{-2mu} 0.5\big(F_{\tilde{L}_n}\big(\tfrac{\thr_B-\Delta}{c_B}\big) \mspace{-2mu} - \mspace{-2mu} F_{\tilde{L}_n}\big(\tfrac{\thr_B+\Delta}{c_B}\big)\big).
\end{align*} 

\vspace{-0.1cm}
\noindent Thus, similarly to the case of system A, the probability of error can be calculated using the standard form of the noise term.

\begin{rem} \label{rem:stableCDFNum}
The work \cite{far15GLOBCOM} does not provide an expression for the CDF $F_{\tilde{L}_n}(\ell_n)$.
However, the CDF of a standardized stable RV can be calculated numerically using the methods described in \cite[Sec. 3]{nol97}. 
\end{rem}

\subsection{System C}
Recall that for system C the two particles are {\em distinguishable}, and $Z_x=T_{x_b}-T_{x_a}$ is the time interval between the releases of particles $b$ and $a$. Here, we assume information is encoded in the {\em order} of release. The system input $Z_x\in\{-\Delta,\Delta\}$ is now given by:
\vspace{-0.15cm}
\begin{align*}
	Z_x = \begin{cases} \Delta, & T_{x_a} = 0, T_{x_b} = \Delta \\
											-\Delta, & T_{x_b} = 0, T_{x_a} = \Delta \end{cases}.
\end{align*}

\vspace{-0.1cm}
\noindent Note that similarly to systems A and B, the information is encoded over the time period $\Delta$.

Let $\tilde{Z}_n \sim (0,1,\tfrac{1}{2},\beta_C)$ be the standard form of the noise term in \eqref{eq:timingChC}. Then the PDF of the system output given the system input is given by
\vspace{-0.15cm}
\begin{align*}
f_{Z_y|Z_x}(z_y|Z_x=-\Delta) &= \frac{f_{\tilde{Z}_n}\big(\tfrac{z_y+\Delta}{c_C}\big)}{c_C}\\
f_{Z_y|Z_x}(z_y|Z_x=\Delta) &= \frac{f_{\tilde{Z}_n}\big(\tfrac{z_y-\Delta}{c_C}\big)}{c_C}. 
\end{align*}

\vspace{-0.1cm}
\noindent Again, to minimize the probability of error at the receiver, the ML detector is used.
Let $\thr_C$ be the optimal ML detection threshold. It is easy to see that this threshold exists since stable distributions are unimodal and the two PDFs are shifted versions of each other. 
The probability of error is now given by: 
\vspace{-0.15cm}
\begin{align}
\label{eq:PeC}
P_e^C &= P(Z_x=-\Delta) \Pr(z_y>\thr_C | Z_x = -\Delta) \nonumber \\
&\qquad +P(Z_x=\Delta) \Pr(z_y\leq\thr_C | Z_x= \Delta), \nonumber \\
&= 0.5 \Pr(z_n>\thr_C+\Delta)+0.5 \Pr(z_n\leq\thr_C-\Delta) \nonumber \\
&= 0.5[1-F_{\tilde{Z}_n}(\tfrac{\thr_C+\Delta}{c_C})+F_{\tilde{Z}_n}(\tfrac{\thr_C-\Delta}{c_C})],
\end{align}

\vspace{-0.1cm}
\noindent which can be calculated using the CDF of the standard form of the noise term. 
Similarly to Remark \ref{rem:stableCDFNum}, the CDF of a standardized RV $\tilde{Z}_n$ can be calculated numerically using the methods described in \cite[Sec. 3]{nol97}. 

\vspace{-0.15cm}
\section{Numerical Evaluation}

\label{sec:numEval}
\vspace{-0.15cm}

We start this section by evaluating the G-SNR definition provided in \eqref{eq:GSNRdef}. In the presence of additive white Gaussian noise, the BER of the ML detector is a function of only SNR, namely, for a fixed SNR the individual values of the signal power and the noise power do not effect the BER.
To show that this property also holds for the considered modulation techniques, we consider system C, which can be specialized to both systems A and B using different values of the parameter $\beta_C$, see \eqref{eq:GSNRa}--\eqref{eq:GSNRc}. Thus, one should expect to observe a constant BER for a fixed value of G-SNR. Note that since $\beta_C$ changes the subclass of the distribution, it must be constant. 

Table~\ref{tab:proofSNR} depicts the BER versus $\Delta$, where the value of G-SNR is kept constant at 1. In this table, the columns corresponds to the values of $\Delta$. 
For each cell, the value of the noise parameter $c_C$ is calculated such that G-SNR is 1. The BER is then numerically calculated using these values based on \eqref{eq:PeC}. It can clearly be observed that the BER is constant for a given G-SNR {\em regardless} of the value of $\Delta$ and $c_C$. It can further be observed that the BER decreases as $\beta_c \rightarrow 1$. This behavior was observed for all three systems for a wide range of G-SNR.

\begin{figure}
		\centering
		\includegraphics[width=0.8\columnwidth]{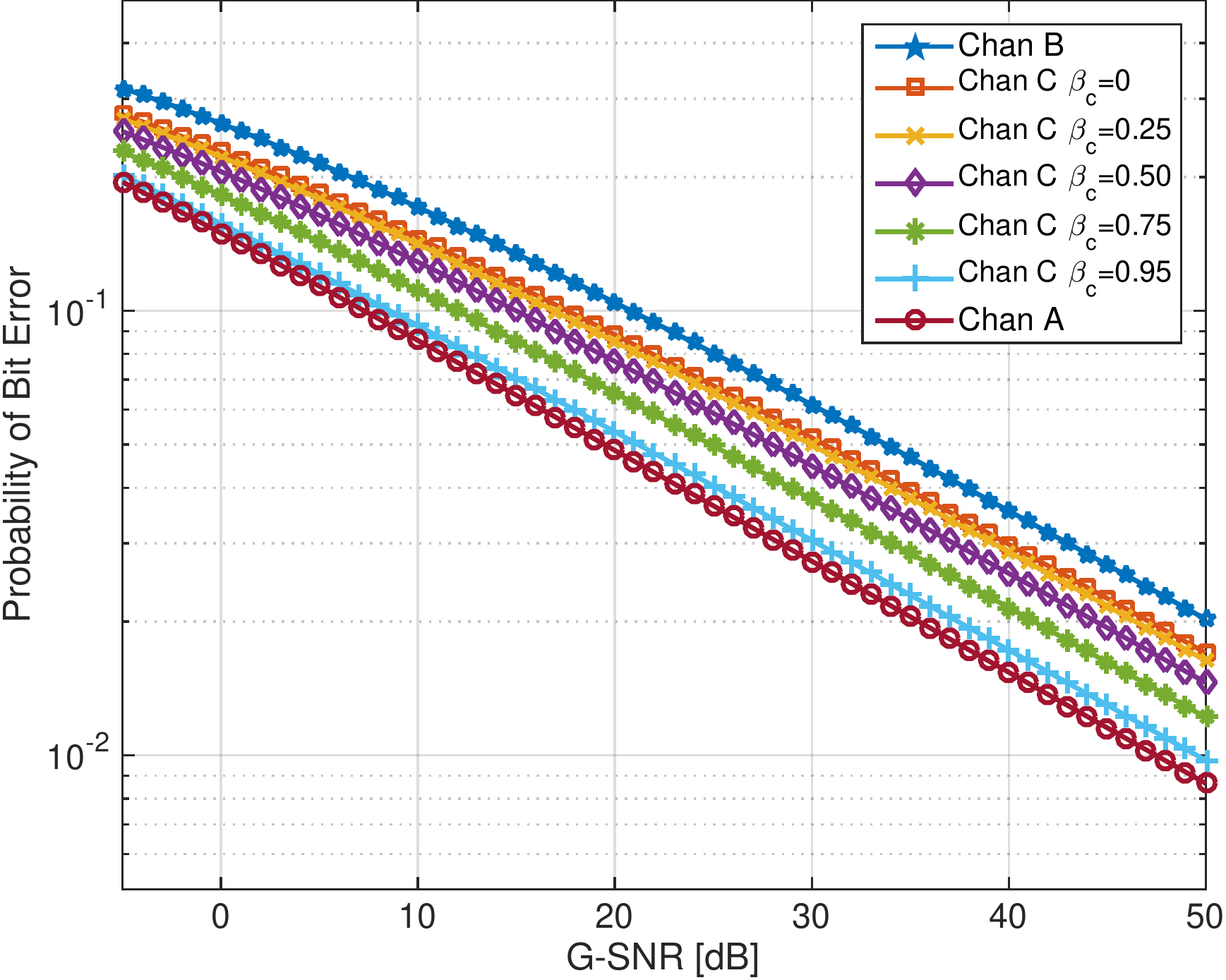}
		\captionsetup{font=small}
		\caption{\label{fig:PBEvsGSNR} BER versus G-SNR in dB for each modulation scheme.}
		\vspace{-0.6cm}
	\end{figure}

Figure~\ref{fig:PBEvsGSNR} depicts the BER versus G-SNR for the different modulation techniques. For system C, five different values of $\beta_C =0,0.25,0.5,0.75,0.95$ are considered. The asynchronous scheme in system B with indistinguishable particles achieves the highest BER, while the system A which assumes perfect synchronization achieves the lowest BER. The gap between these can be thought of as the cost of having no synchronization. Note that in system A, a single particle is released, while in system B two particles are released. This difference will generally entail a higher complexity and cost in System B than in System A.  

\begin{table}[t]
	\begin{center}
				\caption{Constant BER for constant G-SNR. \label{tab:proofSNR}}
		\begin{tabular}[t]{|c|c|c|c|c|}
			\hline
			$\beta$ & $\Delta=0.5$  & $\Delta=5$ & $\Delta=10$ & $\Delta=20$ \\
			\hline
			\hline
			0  &0.1458  &0.1458  &0.1458  &0.1458  \\
			\hline
			0.2 &0.1428  &0.1428  &0.1428  &0.1428  \\
			\hline
			0.5  &0.1287  &0.1287  &0.1287  & 0.1287  \\
			\hline
			0.8 &0.1069  & 0.1069 &0.1069  &0.1069  \\
			\hline
			1  	 &0.0857  & 0.0857 &0.0857  &0.0857 \\
			\hline
		\end{tabular}
	\end{center}
\end{table}

For system C it can be observed that by using two distinguishable particles, the BER improves compared to system~B. Note that when $\beta_C = 0$ the noise distribution is the same as that in system B. In this case, when the dispersion parameter $c$ is the same for both systems, the G-SNR of system C is four times larger than $\gsnr_B^{\text{lb}}$ in \eqref{eq:GSNRb}. Yet, Figure~\ref{fig:PBEvsGSNR} indicates that even for $\beta_C = 0$ the BER of system C is lower than the BER of system B. This demonstrates the destructive effect of the absolute value operation as indicated in Remark \ref{rem:absValImpact}.
Finally, we observe that as $\beta_C$ increases the BER of system~C decreases, while when $\beta_C \rightarrow 1$ the BER of system~C approaches the BER of system A. In this case, asynchronous communication is possible with the same BER performance as synchronized communication at the cost of using two distinguishable particles. 

\vspace{-0.15cm}
\section{Conclusions} \label{sec:conc}
\vspace{-0.15cm}
We have studied the impact of time-synchronization in MT channels by analyzing three modulation techniques. For each of these modulation schemes, and assuming a diffusion-based propagation, the system input and output were related through an additive stable noise. 
Since stable distributions, with the exception of the Gaussian distribution, have infinite variance, we used geometric power as a measure of strength of the noise. Using this approach, we derived the G-SNR for each modulation scheme. Moreover, we characterized the ML detectors for each modulation technique. 
Numerical evaluations show that for a constant G-SNR the BER is constant. This validates the observation that G-SNR, in DBMT channels, plays the role of SNR in additive Gaussian channels.
Finally, we showed that, as expected, synchronization as assumed in the first modulation technique considerably reduces BER relative to other modulation schemes. However, it is possible to achieve a similar BER asynchronously if two distinguishable particles are used per bit, which requires a transmitter that can transmit and a detector that can detect both particles.

\vspace{-0.15cm}
\bibliographystyle{IEEEtran}
\bibliography{IEEEabrv,MolCom_YearSorted}

\end{document}